\documentclass[journal]{IEEEtran}

\usepackage{amsfonts}
\usepackage{amsmath}
\usepackage{mathrsfs}
\usepackage{multirow}
\usepackage{booktabs}
\usepackage{setspace}
\usepackage{citesort}
\usepackage{amsfonts}
\usepackage{amssymb,epic, graphicx}
\usepackage{array}
\usepackage{mdwmath}
\usepackage{mdwtab}
\usepackage{eqparbox}
\usepackage[tight,footnotesize]{subfigure}
\usepackage{algorithm}
\usepackage{algorithmic}
\usepackage{slashbox}
\usepackage{color}

\newtheorem{mythname}{Theorem}

\newtheorem{myexname}{Example}
\newtheorem{myponame}{Proposition}
\newtheorem{mydefname}{Definition}
\usepackage[numbers,sort&compress]{natbib}
\usepackage{graphicx}

\ifCLASSINFOpdf
\else
\fi

\makeatletter
    
    \newcommand{\Rmnum}[1]{\expandafter\@slowromancap\romannumeral #1@}
\makeatother

\begin{document}
%
\title{A New Chase-type Soft-decision Decoding Algorithm for Reed-Solomon Codes}
%
%
%
\author{Siyun~Tang, Suihua~Cai, and Xiao~Ma~\IEEEmembership{Member,
IEEE}

\thanks{Siyun Tang is with the School of Mathematics and System Science, GuangDong Polytechnic Normal University, Guangzhou 510006, China (Email: tangsiy@mail2.sysu.edu.cn).}
\thanks{Suihua Cai is with the School of computer science and engineering, Sun Yat-sen University, Guangzhou 510006, China (Email: caish23@mail.sysu.edu.cn).}
\thanks{Xiao Ma is with the School of computer science and engineering, Sun Yat-sen University, Guangzhou 510006, China (Email: maxiao@mail.sysu.edu.cn).}

\thanks{This work is supported by the Guangdong Provincial NSF under Grants 2018A030313546.}
}

\maketitle

\begin{abstract}
A new Chase-type soft-decision
decoding algorithm for Reed-Solomon codes is proposed, referred to as {\em tree-based
Chase-type algorithm}. The proposed tree-based Chase-type
algorithm takes the set of all vectors as the set of testing
patterns, and hence definitely delivers the most-likely codeword
provided that the computational resources are allowed. All the
testing patterns are arranged in an ordered rooted tree according
to the likelihood bounds of the possibly generated codewords, which is an extension
of Wu and Pados' method from binary into $q$-ary linear block codes.
While performing the algorithm, the ordered rooted tree is
constructed progressively by adding at most two leafs at each
trial. The ordered tree naturally induces a sufficient condition
for the most-likely codeword. That is, whenever the tree-based
Chase-type algorithm exits before a preset maximum number of
trials is reached, the output codeword must be the most-likely
one. But, in fact, the algorithm can be terminated by setting a
discrepancy threshold instead of a maximum number of trials.
When the tree-based Chase-type algorithm is combined with
Guruswami-Sudan (GS) algorithm, each trial can be implement in an
extremely simple way by removing from the gradually updated
Gr\"{o}bner basis one old point and interpolating one new point.
Simulation results show that the tree-based Chase-type algorithm
performs better than the recently proposed Chase-type algorithm by
Bellorado et al with less trials~(on average) given that the
maximum number of trials is the same.

\end{abstract}

\begin{IEEEkeywords}
Error-correction codes, Chase-type algorithm, flipping patterns, Guruswami-Sudan
algorithm, hard-decision deocoding, Reed-Solomon codes, soft-decision decoding.
\end{IEEEkeywords}

\IEEEpeerreviewmaketitle

\section{Introduction}\label{introduction}
\IEEEPARstart{R}{eed-Solomon} (RS) codes are an important class of
algebraic codes, which have been widely used in many practical
systems, including space and satellite communications, data
storage, digital audio/video transmission and file
transfer~\cite{costello1998applications}. Even more contemporary codes such as
turbo product codes~(TPC)~\cite{Al-Dweik09btc,Al-Dweik09tpc,Al-Dweik11tpc}
can be used with RS codes. The widespread use of RS
codes is primarily due to their large error-correction capability,
a consequence of their maximum distance separable~(MDS) property.
Investigating the decoding algorithms for RS codes is important
both in practice and in theory. The traditional hard-decision
decoding~(HDD) algorithms, such as Berlekamp-Massey
(BM)~\cite{Berlekamp68'}, Welch-Berlekamp (WB)~\cite{Welch83} and
Euclidean~\cite{Sugiyama75} algorithms, are efficient to find the
unique codeword~(if exists) within a Hamming sphere of radius less
than the half minimum Hamming distance. Hence, their
error-correction capability is limited by the half minimum Hamming
distance bound. In contrast, Guruswami-Sudan~(GS)
algorithm~\cite{Sudan97}\cite{Guruswami99} can enlarge the
decoding radius and may output a list of candidate codewords.
Hence, GS algorithm can correct errors beyond the half minimum
Hamming distance bound. Nevertheless, HDD algorithms do not fully exploit the information provided by the channel output. The Koetter-Vardy~(KV) algorithm~\cite{Koetter03}
extended the GS decoder to an algebraic soft-decision~(ASD) decoding
algorithm by transforming the soft information into the multiplicity matrix
that is then taken as input to the GS decoder.

The KV algorithm outperforms the
GS algorithm but suffers from high complexity. To reduce the
complexity, a progressive list-enlarged algebraic soft-decision decoding
algorithm has been proposed
in~\cite{tang2012progressive}\cite{tang2013progressive}.
The other class of soft-decision decoding~(SDD) algorithms,
which are based on using multiple trials of a low-complexity RS
decoding algorithm in combination with erasing or flipping a set of symbols or bits in each trial, have renewed the interest of researchers. These algorithms use the soft-information available at the channel output to construct a set of either erasure patterns~\cite{Forney66,Kotter96,lee2008soft,Nguyen11}, flipping patterns~\cite{Chase72,wu2003adaptive,Bellorado10}, or patterns combining both~\cite{Tang01}\cite{Xia08}, and then attempt to decode using each pattern. Meanwhile, more related works focus on how to select patterns~\cite{valls2019,jeong2021}, how to rule out useless patterns~\cite{koumoto1998sufficient} and how to identify the maximum-likelihood~(ML) codeword~\cite{taipale1991improvement}\cite{Kaneko94}.
Other soft-decision decoding algorithms based on multiple trials can be
found in~\cite{fossorier1995soft,jiang2006iterative,Bellorado10Ping}.
In~\cite{fossorier1995soft}, re-encoding is performed for each
trial, where the generator matrix is adapted based on most
reliable positions~(MRPs) specified by the ordered statistics.
In~\cite{jiang2006iterative} and~\cite{Bellorado10Ping}, a decoding algorithm combined with belief propagation is performed for each trial, where the
parity-check matrix is iteratively adapted based on the LRPs.

In this paper, we focus on the Chase-type decoding algorithm. Let
$\mathcal{C}_q[n, k]$ be an RS code over the finite field of size
$q$ with length $n$, dimension $k$, and the minimum Hamming
distance $d_{\min} = n - k + 1$, Generally, a Chase-type
soft-decision decoding algorithm has three ingredients: 1) a set
of {\em flipping patterns} $\mathcal{F}
\stackrel{\Delta}{=}\{\textbf{f}^{(0)},\cdots,
\textbf{f}^{(L-1)}\}$ where $\textbf{f}^{(\ell)}$ is a vector of
length $n$, 2) a hard-decision decoder~(HDD), and 3) a stopping
criterion. Given these three ingredients, a Chase-type decoding
algorithm works as follows. For each $\ell \geq 0$, the Chase-type
algorithm makes a trial by decoding $\textbf{z} -
\textbf{f}^{(\ell)}$ with the HDD. If the HDD is successful, the
output is referred to as a candidate codeword. Once some candidate
is found to satisfy the stopping criterion, the algorithm
terminates. Otherwise, the algorithm chooses as output the most
likely candidate after all flipping patterns in $\mathcal{F}$ are
tested.

The Chase-2 decoding algorithm~\cite{Chase72}, which is well known for binary codes, finds the $t_{\min}\stackrel{\Delta}{=} \lfloor (d_{\min}-1)/2 \rfloor$ LRPs, and then constructs the commonly-used $\mathcal{F}$ by $2^{t_{\min}}$ flipping patterns combinatorially. Obviously, the straightforward generalization of Chase-2 from binary to nonbinary incurs high complexity especially for large $q$ and $d_{\min}$. Later, Wu and Pados~\cite{wu2003adaptive} proposed an adaptive two-stage algorithm for ML/sub-ML decoding, where in the first stage the flipping patterns in $\mathcal{F}$ are constructed dynamically according to their ascending lower bound on the soft discrepancy. With this testing order, the ML codeword is expected to appear earlier. This order also provides a sufficient condition to terminate the algorithm before all flipping patterns are tested. However, Wu and Pados did not show explicitly in~\cite{wu2003adaptive} the way to generate the ordered flipping patterns, especially for nonbinary codes. Next, a low-complexity Chase~(LCC) decoding algorithm~\cite{Bellorado10} for RS codes, which can achieve better performance-complexity tradeoff, constructs $2^{\eta}$ flipping patterns by first finding $\eta$ LRPs and then restricts only two most likely symbols at each LRP. For the HDD algorithm required in the Chase-type algorithm, one usually chooses a simple HDD algorithm, such as BM algorithm. In contrast, the LCC algorithm implements the GS decoding algorithm~(with multiplicity one) for each trial. This algorithm has a clear advantage when two flipping patterns diverge in only one coordinate, in which case backward interpolation architecture~\cite{zhang2013generalized} and the module basis reduction interpolation technique~\cite{xing2020} can be
employed to further reduce the decoding complexity. Because of this, even though the set of erasure patterns designed using rate-distortion theory in~\cite{Nguyen11} may slightly outperform the LCC algorithm with the same number of decoding trials, the latter may require lower implementation complexity.

In this paper, we propose to arrange all possible flipping patterns into
an ordered rooted tree, which is constructed progressively by adding
at most two leafs at each trial. With the aid of the inherent structure
of the tree, we devise an efficient algorithm that generates one-by-one upon request flipping patterns according to their soft discrepancy lower bounds. To expedite the decoding, we propose a sufficient condition that generalizes Taiple
and Pursley's condition~\cite{Taipale94} to prune some unnecessary sub-trees. In addition, when the new algorithm is combined with the GS algorithm, each trial can be implement in an extremely simple way by removing from the gradually updated Gr\"{o}bner basis one old point and interpolating one new point. Simulation results show that the proposed algorithm performs better than the LCC algorithm~\cite{Bellorado10} with less trials~(on average) given that the maximum number of trials is the same. To illustrate the near-optimality of the proposed algorithm in the high
signal-to-noise ratio~(SNR) region, we also propose a method to simulate performance bounds on the maximum-likelihood decoding~(MLD) algorithm.


The rest of this paper is organized as follows.
Sec.~\ref{sec:TESTORDER}~defines the ordered rooted tree of
flipping patterns and provides a general framework of the
tree-based Chase-type algorithm. In Sec.~\ref{sec:TCGA}, the
tree-based Chase-type algorithm is combined with the GS algorithm.
Numerical results and further discussion are presented in
Sec.~\ref{sec:PACC}.
Sec.~\ref{sec:Conclusions} concludes this paper.

\section{Testing Order of Flipping Patterns}\label{sec:TESTORDER}
\subsection{Basics of RS Codes}
Let $\mathbb{F}_q \stackrel{\Delta}{=}
\{\alpha_0,\alpha_1,\cdots,\alpha_{q-1}\}$ be the finite field of
size $q$. A codeword of the RS code $\mathcal{C}_q[n, k]$ can be
obtained by evaluating a polynomial of degree less than $k$ over
$n$ distinct points, denoted by $\mathcal {L} \stackrel{\Delta}{=}
\{\beta_0, \beta_1, \cdots, \beta_{n-1}\} \subseteq \mathbb{F}_q$.
To be precise, the codeword corresponding to a message polynomial
$u(x) = u_0 + u_1x +\cdots+ u_{k-1}x^{k-1}$ is given by
\begin{equation*}
    \textbf{c}\! =\!\! (c_0,\! c_1,\!\cdots,\! c_{n-1})\! =\! (u(\beta_0),\! u(\beta_1),\! \cdots,\!u(\beta_{n-1})).
\end{equation*}
Assume that the codeword $\textbf{c}$ is transmitted through a
memoryless channel, resulting in a received vector
\begin{equation*}
    \textbf{r}= (r_0, r_1,\cdots, r_{n-1}).
\end{equation*}
The corresponding hard-decision vector is denoted by
\begin{equation*}
    \textbf{z}= (z_0, z_1,\cdots, z_{n-1}),
\end{equation*}
where $z_j \stackrel{\Delta}{=} \arg \max_{\alpha \in
\mathbb{F}_q} {\rm Pr}(r_j|\alpha), 0 \leq j \leq n-1$. Here, we are
primarily concerned with additive white Gaussian noise~(AWGN)
channels. In this scenario, a codeword is modulated into a
real signal before transmission and the channel transition
probability function ${\rm Pr}(r_j|\alpha)$ is replaced by the
conditional probability density function.

The {\em error pattern} is defined by $\textbf{e}
\stackrel{\Delta}{=} \textbf{z} - \textbf{c}$. A conventional
hard-decision decoder~(HDD) can be implemented to find the
transmitted codeword whenever the Hamming weight $W_H(\textbf{e})$
is less than or equal to $t_{\min} \stackrel{\Delta}{=}
\lfloor(n-k)/2\rfloor$. The HDD is simple, but it usually causes
performance degradation. In particular, it even fails to output a
valid codeword if $\textbf{z}$ lies at Hamming distance greater
than $t_{\min}$ from any codeword. An optimal decoding
algorithm~(to minimize the word error probability when every
codeword is transmitted equal-likely) is the {\em maximum
likelihood decoding} (MLD) algorithm, which delivers as output the
codeword $\textbf{c}$ that maximizes the log-likelihood metric
$\sum_{j=0}^{n-1}\log {\rm Pr}(r_j|c_j)$. The MLD algorithm is
able to decode beyond $t_{\min}$ errors, however, it is
computationally infeasible in general~\cite{Guruswami2005maximum}. A Chase-type soft-decision
decoding algorithm trades off between the HDD and the MLD by
performing the HDD successively on a set of flipping patterns.

\subsection{Minimal Decomposition of Hypothesized Error Patterns}
Similar to~\cite[Chapter 3]{gallager1963}, we introduce the following definitions.
\begin{mydefname}
Let $\textbf{z}$ be the hard-decision vector. A {\em hypothesized error pattern} $\textbf{e}$
is defined as a vector such that $\textbf{z} - \textbf{e}$ is a valid codeword.
\hfill{$\square$}
\end{mydefname}

Notice that $\textbf{z}$ itself is a hypothesized error pattern
since $\textbf{z} - \textbf{z}$ is the all-zero codeword. To each
component $e_j = z_j - c_j $ of the hypothesized error pattern, we
assign a {\em soft weight} $\lambda_j(e_j)\stackrel{\Delta}{=}
\log {\rm Pr}(r_j|z_j) - \log {\rm Pr}(r_j|c_j)$. The soft weight
of a hypothesized error pattern $\textbf{e}$ is defined as
$\lambda(\textbf{e}) = \sum_{j=0}^{n-1} \lambda_j(e_j) = \sum_{j:
e_j \neq 0} \lambda_j(e_j)$. The MLD algorithm can be equivalently
described as finding one {\em lightest} hypothesized error pattern
$\textbf{e}^*$ that minimizes $\lambda(\textbf{e})$.

Since the soft weight of a hypothesized
error pattern $\textbf{e}$ is completely determined by its nonzero components, we
may simply list all its non-zero components. For clarity, a nonzero component $e_j$ of $\textbf{e}$ is denoted by $(j, \delta)$ meaning that an error of value $\delta$ occurs at the $j$-th coordinate,
i.e., $e_j = \delta$. For convenience, we call $(j, \delta)$ with $\delta \neq 0$ an {\em atom}. In the following, we will define a total order over the set of all $n(q-1)$ atoms. For the purpose of tie-breaking, we define simply a
total order over the field $\mathbb{F}_q$ as $\alpha_0 < \alpha_1
< \cdots < \alpha_{q-1}$.

\begin{mydefname}
We say that $(j, \delta) \prec (j',
\delta')$ if and only if $\lambda_j(\delta) <
\lambda_{j'}(\delta')$ or $\lambda_j(\delta) =
\lambda_{j'}(\delta')$ and $j < j'$ or $\lambda_j(\delta) =
\lambda_{j'}(\delta')$, $j = j'$ and $\delta < \delta'$.
\end{mydefname}

With this definition, we can arrange all the $n(q-1)$ atoms into a
chain~(denoted by $\textbf{A}$ and referred to as {\em atom chain}) according to the increasing order. That is,
\begin{equation}\label{total_order}
  {\textbf A}\! \stackrel{\Delta}{=}\! [(j_1,\! \delta_1)\! \prec \! (j_2,\! \delta_2)\! \prec
 \! \cdots \! \prec \! (j_{n(q-1)},\! \delta_{n(q-1)})].
\end{equation}

The rank of an atom $(j, \delta)$, denoted by $Rank(j, \delta)$,
is defined as its position in the atom chain $\textbf{A}$.

\begin{mydefname}
Let $\textbf{f}$  be a nonzero vector. Its {\em
support set} is defined as $\mathcal{S}(\textbf{f})
\stackrel{\Delta}{=}\{j: f_j \neq 0 \}$, whose cardinality
$|\mathcal{S}(\textbf{f})|$ is the Hamming weight
$W_H(\textbf{f})$ of $\textbf{f}$. Its {\em lower rank} and {\em upper rank} are
defined as $R_{\ell}(\textbf{f}) \stackrel{\Delta}{=}\min_{f_j
\neq 0} Rank(j, f_j)$ and $R_{u}(\textbf{f})
\stackrel{\Delta}{=}\max_{f_j \neq 0} Rank(j, f_j)$, respectively.
\hfill{$\square$}
\end{mydefname}

We assume that $R_{\ell}(\textbf{0}) = +\infty$ and  $R_{u}(\textbf{0}) = -\infty$.

\begin{myponame}\label{fpattern}
Any nonzero vector $\textbf{f}$ can be represented in a unique way by listing all
its nonzero components as
\begin{equation}\label{incre_err}
  {\textbf f}\stackrel{\Delta}{=}[(i_1, \gamma_1) \prec (i_2, \gamma_2)  \prec \cdots \prec (i_t, \gamma_t)],
\end{equation}
where $t = W_H(\textbf{f})$, $R_{\ell}(\textbf{f}) = Rank(i_1, \gamma_1)$ and $R_{u}(\textbf{f}) = Rank(i_t, \gamma_t)$.
\end{myponame}

\begin{IEEEproof}
It is obvious and omitted here.
\end{IEEEproof}

Proposition~\ref{fpattern} states that any nonzero vector can be
viewed as a sub-chain of $\textbf{A}$. In contrast, any sub-chain
of $\textbf{A}$ specifies a nonzero vector {\em only when} all
atoms in the sub-chain have distinct coordinates.

\begin{myponame}\label{edecompose}
Any nonzero vector $\textbf{e}$ with
$W_H(\textbf{e}) \geq t_{\min}$ can be {\em uniquely} decomposed as
$\textbf{e} = \textbf{f} + \textbf{g}$ satisfying that $|\mathcal{S}(\textbf{g})| = t_{\min}$, $\mathcal{S}(\textbf{f}) \cap \mathcal{S}(\textbf{g}) = \emptyset$ and $R_u(\textbf{f}) < R_\ell(\textbf{g})$.
\end{myponame}

\begin{IEEEproof}
From Proposition~\ref{fpattern}, we have $\textbf{e}  = [(i_1, \gamma_1)\prec \cdots \prec (i_t, \gamma_t)]$ where $t = W_H(\textbf{e})$. Then this proposition can be verified by defining $\textbf{f} = [(i_1, \gamma_1)\prec \cdots \prec (i_{t-t_{\min}}, \gamma_{t-t_{\min}})]$ and $\textbf{g} = [(i_{t-t_{\min}+1}, \gamma_{t-t_{\min}+1})\prec \cdots \prec (i_t, \gamma_t)]$.
\end{IEEEproof}

For a vector $\textbf{f}$, define $\mathcal{G}(\textbf{f}) =
\{\textbf{g}: |\mathcal{S}(\textbf{g})| = t_{\min},
\mathcal{S}(\textbf{f}) \cap \mathcal{S}(\textbf{g}) = \emptyset,
R_u(\textbf{f}) < R_\ell(\textbf{g})\}$. From
Proposition~\ref{edecompose}, any hypothesized error pattern
$\textbf{e}$ with $W_H(\textbf{e}) \geq t_{\min}$ can be
decomposed as $\textbf{e} = \textbf{f} + \textbf{g}$ in a unique
way such that $\textbf{g} \in \mathcal{G}(\textbf{f})$. This
decomposition is referred to as the {\em minimal
decomposition}\footnote{This terminology comes from the fact as
shown in Appendix~\ref{sec:Minimalityoffp}.}, where $\textbf{f}$ is referred to as
the {\em minimal flipping pattern} associated with $\textbf{e}$.
In the case when a hypothesized error pattern $\textbf{e}$ exists
with $W_H(\textbf{e}) < t_{\min}$, we define $\textbf{0}$ as the
{\em minimal flipping pattern} associated with $\textbf{e}$.

For every $\textbf{f} \in \mathbb{F}_q^n$, when taking $\textbf{z}
- \textbf{f}$ as an input vector, the HDD either reports a
decoding failure or outputs a unique codeword $\textbf{c}$. In the
latter case, we say that the flipping pattern $\textbf{f}$ {\em
generates} the hypothesized error pattern $\textbf{e} = \textbf{z}
- \textbf{c}$. Actually, any hypothesized error pattern can be generated by many flipping patterns (collectively referred to as equivalence class in~\cite{moorthy1997soft}). Apparently, an efficient algorithm should avoid to test multiple flipping patterns in the same equivalence class. As pointed out in~\cite{moorthy1997soft}\cite{koumoto1998sufficient}, we have the following proposition.

\begin{myponame}\label{egeneratedbyf}
Any hypothesized error pattern can be generated by its associated minimal flipping pattern.
\end{myponame}
\begin{IEEEproof}
It is obvious.
\end{IEEEproof}

From Proposition~\ref{egeneratedbyf}, in principle, we only need
to decode all vectors $\textbf{z} - \textbf{f}$ with the minimal
flipping patterns $\textbf{f}$. Unfortunately, we do not know
which flipping patterns are minimal before performing the HDD.
Even worse, we do not know whether or not a vector $\textbf{f}$
can generate a hypothesized error pattern before performing the
HDD. However, by extending the Proposition~1 in~\cite{wu2003adaptive}, we have the following theorem, which provides a lower bound on the soft weight of the generated error pattern whenever $\textbf{f}$ is a minimal flipping pattern.

\begin{mythname}\label{elowerbound}
Let $\textbf{f}$ be a nonzero vector that is the minimal flipping pattern to generate a hypothesized error pattern $\textbf{e}$.
Then $\lambda(\textbf{e}) \geq \lambda(\textbf{f}) + \min_{\textbf{g}\in \mathcal{G}(\textbf{f})} \lambda(\textbf{g})$.
\end{mythname}

\begin{IEEEproof}
For $W_H(\textbf{e}) > t_{\min}$, from
Proposition~\ref{edecompose}, we have the minimal decomposition
$\textbf{e} = \textbf{f} + \textbf{g}$, $\textbf{g} \in
\mathcal{G}(\textbf{f})$. Hence, $\lambda(\textbf{e}) =
\lambda(\textbf{f}) + \lambda(\textbf{g}) \geq \lambda(\textbf{f})
+ \min_{\textbf{g}\in
\mathcal{G}(\textbf{f})}\lambda(\textbf{g})$.

\end{IEEEproof}

More importantly, the lower bound given in
Theorem~\ref{elowerbound} is computable for any nonzero vector
$\textbf{f}$ without performing the HDD since $\min_{\textbf{g}
\in \mathcal{G}(\textbf{f})} \lambda(\textbf{g})$ can be
calculated using the following greedy algorithm with the help of the atom chain $\textbf{A}$.

{\em \textbf{Algorithm} 1:} Greedy Algorithm for Computing
$\min_{\textbf{g}\in \mathcal{G}(\textbf{f})}\lambda(\textbf{g})$.
\begin{itemize}
  \item {\bf Input:} A nonzero vector $\textbf{f}$.
  \item {\bf {Initialization:}} Set $\textbf{g} = \textbf{0}$, $\lambda(\textbf{g}) = 0$, $W_H(\textbf{g}) = 0$ and $i = R_u(\textbf{f}) + 1$.
  \item {\bf {Iterations:}} While $W_H(\textbf{g}) < t_{\min}$ and $i \leq n(q-1)$, do
\begin{enumerate}
  \item if $j_i \notin \mathcal{S}(\textbf{f}+\textbf{g})$, let~$\lambda(\textbf{g}) \leftarrow \lambda(\textbf{g}) + \lambda_{j_i}(\delta_i)$, $\textbf{g} \leftarrow \textbf{g} + (j_i, \delta_i)$ and $W_H(\textbf{g}) \leftarrow W_H(\textbf{g}) + 1$.
  \item $i \leftarrow i + 1$.
\end{enumerate}
  \item {\bf Output:} If $W_H(\textbf{g}) = t_{\min}$, output $\min_{\textbf{g}\in \mathcal{G}(\textbf{f})}\lambda(\textbf{g}) = \lambda(\textbf{g})$.
   Otherwise, we must have $\mathcal{G}(\textbf{f}) = \emptyset$; in this case, output $\min_{\textbf{g}\in \mathcal{G}(\textbf{f})}\lambda(\textbf{g}) = +\infty$.
\end{itemize}

The correctness of the above greedy algorithm can be argued as
follows. Let $\textbf{g}^*$ be the sub-chain of $\textbf{A}$ found
when the algorithm terminates. This sub-chain must be a vector
since no two atoms contained in $\textbf{g}^*$ can have the same
coordinate for that each atom $(j_i, \delta_i)$ is added only when
$j_i \notin \mathcal{S}(\textbf{f}+\textbf{g})$.
We only need to consider the case when $\mathcal{G}(\textbf{f}) \neq \emptyset$, which is equivalent to saying that all atoms with rank greater than $R_u(\textbf{f})$ occupy at least $t_{\min}$ coordinates.
In this case, we must have $W_H(\textbf{g}^*) = t_{\min}$. We then
have $\textbf{g}^* \in \mathcal{G}(\textbf{f})$ since
$\mathcal{S}(\textbf{f}) \cap \mathcal{S}(\textbf{g}^*) =
\emptyset$ and $R_u(\textbf{f}) < R_{\ell}(\textbf{g}^*)$ for
that $i$ begins with $R_u(\textbf{f}) + 1$ and each atom $(j_i,
\delta_i)$ is added only when $j_i \notin
\mathcal{S}(\textbf{f}+\textbf{g})$. The minimality of $\lambda(\textbf{g}^*)$ can be proved by induction on the iterations.

\subsection{Tree of Flipping Patterns}
All flipping patterns are arranged in an ordered rooted tree, denoted by $\textbf{T}$, as described below.
\begin{itemize}
  \item[T1.] The root of the tree is $\textbf{f} = \textbf{0}$, which is located at the 0-th level. For $i\geq 1$, the $i$-th level of the tree consists of all nonzero vectors with Hamming weight $i$.
  \item[T2.] A vertex $\textbf{f}$ at the $i$-th level takes as children all vectors from $\{\textbf{f} + (j, \delta): (j, \delta)\in \textbf{A}, Rank(j, \delta) > R_{u}(\textbf{f}), j \notin \mathcal{S}(\textbf{f})\}$, which are arranged at the $(i+1)$-th level from left to right with increasing upper ranks. The root has $n(q-1)$ children. A nonzero vertex $\textbf{f}$ has at most $n(q-1) - R_{u}(\textbf{f})$ children.
\end{itemize}

For each vertex $\textbf{f}$ in $\textbf{T}$, define
\begin{equation}\label{Bound1}
  B(\textbf{f}) \stackrel{\Delta}{=} \left\{\begin{array}{cc}
                 \lambda(\textbf{f}) + \min_{\textbf{g} \in \mathcal{G}(\textbf{f})} \lambda(\textbf{g}), & if~\mathcal{G}(\textbf{f}) \neq \emptyset; \\
                 +\infty, & otherwise
               \end{array}\right..
\end{equation}

\begin{mythname}\label{VertexonTree}
Let $\textbf{f}$ be a vertex. If exist, let $\textbf{f}_\uparrow$ be its parent, $\textbf{f}_\downarrow$ be one of its children and $\textbf{f}_\rightarrow$ be one of its right-siblings.  We have $B(\textbf{f}) \leq B(\textbf{f}_\downarrow)$ and $B(\textbf{f}) \leq B(\textbf{f}_\rightarrow)$.
\end{mythname}

\begin{IEEEproof}
We have $\lambda(\textbf{f}_\downarrow) > \lambda(\textbf{f})$ since $\textbf{f}_\downarrow$ has one more atom than $\textbf{f}$. We also have $\min_{\textbf{g} \in \mathcal{G}(\textbf{f}_\downarrow)} \lambda(\textbf{g}) \geq \min_{\textbf{g} \in \mathcal{G}(\textbf{f})} \lambda(\textbf{g}) $ since $\mathcal{G}(\textbf{f}_\downarrow) \subseteq \mathcal{G}(\textbf{f})$. Therefore, $B(\textbf{f}_\uparrow) \leq B(\textbf{f}) \leq B(\textbf{f}_\downarrow)$.

For a nonzero vertex $\textbf{f}$, we have $\textbf{f} = \textbf{f}_\uparrow + (j, \delta)$ and $\textbf{f}_\rightarrow = \textbf{f}_\uparrow + (j', \delta')$ where $Rank(j, \delta) < Rank(j', \delta')$. Hence $\lambda_j(\delta) \leq \lambda_{j'}(\delta')$, which implies that $\lambda(\textbf{f}) \leq \lambda(\textbf{f}_\rightarrow)$.
Let $\textbf{g}_\rightarrow \in \mathcal{G}(\textbf{f}_\rightarrow)$ such that $\lambda(\textbf{g}_\rightarrow)  = \min_{\textbf{g}\in \mathcal{G}(\textbf{f}_\rightarrow)} \lambda(\textbf{g})$. If $j \in \mathcal{S}(\textbf{g}_\rightarrow)$, define $\textbf{g} = \textbf{g}_\rightarrow - (j, \delta'') + (j', \delta')$, where $(j, \delta'')$ is an atom contained in $\textbf{g}_\rightarrow$; otherwise, define $\textbf{g} = \textbf{g}_\rightarrow$. In either case, we can verify that $\textbf{g} \in \mathcal{G}(\textbf{f})$ and $\lambda(\textbf{g}) \leq  \lambda(\textbf{g}_\rightarrow)$, implying that $\min_{\textbf{g} \in \mathcal{G}(\textbf{f})}\lambda(\textbf{g}) \leq \min_{\textbf{g} \in \mathcal{G}(\textbf{f}_\rightarrow)}\lambda(\textbf{g})$. Therefore, we have $B(\textbf{f}) \leq B(\textbf{f}_\rightarrow)$.
\end{IEEEproof}

\begin{mydefname}
A subtree $\textbf{T}'$ is said to be {\em sufficient} for the MLD algorithm if the lightest hypothesized error pattern can be generated by some vertex in $\textbf{T}'$.
\end{mydefname}

By this definition, we can see that $\textbf{T}$ is itself sufficient. We can also see that removing all vertexes with Hamming weight greater than $n-t_{\min}$ does not affect the sufficiency. Generally, we have
\begin{mythname}\label{removebound}
Let $\textbf{e}^*$ be an available hypothesized error pattern and $\textbf{f}$ be a nonzero vertex such that $B(\textbf{f}) \geq \lambda(\textbf{e}^*)$. Then removing the subtree rooted from $\textbf{f}$ does not affect the sufficiency.
\end{mythname}

\begin{IEEEproof}
If exists, let $\textbf{e}$ be a hypothesized error pattern such that $\lambda(\textbf{e}) < \lambda(\textbf{e}^*)$.
It suffices to prove that $\textbf{e}$ can be generated by some vertex in the remaining subtree. Let $\textbf{h}$ be the minimal flipping pattern associated with $\textbf{e}$.
From Theorem~\ref{elowerbound}, we have $B(\textbf{h}) \leq \lambda(\textbf{e}) <  \lambda(\textbf{e}^*) \leq B(\textbf{f})$. From Theorem~\ref{VertexonTree}, $\textbf{h}$ is not contained in the subtree rooted from $\textbf{f}$ and hence has not been removed.
\end{IEEEproof}

Similar to~\cite{wu2003adaptive}, a total order of all vertexes can be defined as follows.
\begin{mydefname}
We say that $\textbf{f} \prec \textbf{h}$ if and only if $B(\textbf{f}) < B(\textbf{h})$ or $B(\textbf{f}) = B(\textbf{h})$ and $W_H(\textbf{f}) < W_H(\textbf{h})$ or $B(\textbf{f}) = B(\textbf{h})$, $W_H(\textbf{f}) = W_H(\textbf{h})$ and $\textbf{f}$ is located at the left of $\textbf{h}$.
\end{mydefname}

Suppose that we have an efficient algorithm that can generate {\em one-by-one upon request} all flipping patterns in the following order
\begin{equation}\label{B1Order}
  \mathcal{F} \stackrel{\Delta}{=} \textbf{f}^{(0)} \prec \textbf{f}^{(1)} \prec \cdots \prec \textbf{f}^{(i)} \prec \cdots
\end{equation}
Then we can perform a Chase-type algorithm as follows. For $i = 0, 1, \cdots $, we perform the HDD by taking $\textbf{z} - \textbf{f}^{(i)}$ as the input. If a hypothesized error pattern $\textbf{e}^*$ is found satisfying $\lambda(\textbf{e}^*) \leq B(\textbf{f}^{(i)})$, then the algorithm terminates. Otherwise, the process continues until a preset maximum number of trials is reached. The structure of the tree $\textbf{T}$ is critical to design such an efficient sorting algorithm. From Theorem~\ref{VertexonTree}, we know that $\textbf{f} \prec \textbf{f}_\downarrow$ and $\textbf{f} \prec \textbf{f}_\rightarrow$.  Therefore, $\textbf{f}^{(i)}$ must be either the left-most child of some $\textbf{f}^{(j)}$ with $j < i$ or the following~(adjacent) right-sibling of some $\textbf{f}^{(j)}$ with $j < i$. In other words, it is not necessary to consider a flipping pattern $\textbf{f}$ before both its parent and its preceding left-sibling are tested. This motivates the following Chase-type algorithm, referred to as {\em tree-based Chase-type algorithm}.

{\em \textbf{Algorithm} 2:} A General Framework of Tree-based Chase-type Algorithm
\begin{itemize}
  \item {\bf Preprocessing:} Find the hard-decision vector $\textbf{z}$; calculate the soft weights of $n(q-1)$ atoms; construct the atom chain $\textbf{A}$. Suppose that we have a {\em linked list} $\mathcal{F} = \textbf{f}^{(0)} \prec \textbf{f}^{(1)} \prec \textbf{f}^{(2)} \prec \cdots$, which is of size at most $L$ and maintained in order during the iterations.
  \item {\bf Initialization:} $\mathcal{F} = \textbf{0}$; $\ell = 0$; $\textbf{e}^* = \textbf{z}$.
  \item {\bf Iterations:} While $\ell < L$, do the following.
      \begin{enumerate}
        \item If $\lambda(\textbf{e}^*) \leq B(\textbf{f}^{(\ell)})$, output $\textbf{e}^*$ and exit the algorithm;
        \item Perform the HDD by taking $\textbf{z} - \textbf{f}^{(\ell)}$ as input;
        \item In the case when the HDD outputs a hypothesized error pattern $\textbf{e}$ such that $\lambda(\textbf{e})  < \lambda(\textbf{e}^*)$, set $\textbf{e}^* = \textbf{e}$;
        \item Update the linked list $\mathcal{F}$ by inserting~(if exist) the left-most child and the following right-sibling of $\textbf{f}^{(\ell)}$ and removing $\textbf{f}^{(\ell)}$;
        \item Increment $\ell$ by one.
  \end{enumerate}
\end{itemize}

\textbf{Remarks.}
\begin{itemize}
  \item The vector $\textbf{e}^*$ can be initialized by any hypothesized error pattern~(say obtained by the re-encoding approach) other than $\textbf{z}$. Or, we may leave $\textbf{e}^*$ uninitialized and   set $\lambda(\textbf{e}^*) = +\infty$ initially.
  \item
    Obviously, the computational complexity of preprocessing step is $O(n(q-1)\log(n(q-1)))$. Note that, in each iteration of Algorithm 2, one flipping pattern is removed from $\mathcal{F}$ and at most two flipping patterns are inserted into $\mathcal{F}$. Also note that the size of $\mathcal{F}$ can be kept as less than or equal to $L-\ell$ by removing extra tailed flipping patterns. Hence maintaining the linked list $\mathcal{F}$ requires computational complexity of order at most $O(L\log L)$, which is consistent with~\cite[Theorem 1]{wu2003adaptive}. If the algorithm exits within $L$ iterations, the found hypothesized error pattern must be the lightest one.
  \item
    The above algorithm is an extension of this process, which is referred to as LB ascending search in~\cite{wu2003adaptive} and that is the optimal searching order in the sense that the HDD approaches rapidly the ML codeword, i.e., the lightest error pattern. During the iterations, sufficient conditions in~\cite{taipale1991improvement} or Lemma~1 in~\cite{Kaneko94} for $q$-ary RS codes~(see the corrected form in~\cite{tang2012correction})
    can also be used to identify the lightest error pattern. The lemma is rephrased as follows.
\end{itemize}

\begin{mythname}\label{Kanekolemma} If an error pattern $\textbf{e}^*$ satisfies

\begin{equation}\label{Kanekobound}
 \lambda(\textbf{e}^*) \leq B_0(\textbf{e}^*) \stackrel{\Delta}{=}\min_{\textbf{e}\in \mathcal{E}(\textbf{e}^*)} \lambda(\textbf{e}),
\end{equation}
where $\mathcal{E}({\textbf e}^*) = \{\textbf{e}: \mathcal{S}(\textbf{e}^*) \cap \mathcal{S}(\textbf{e}) = \emptyset$, $|\mathcal{S}(\textbf{e})| = d_{\min} - W_H(\textbf{e}^*)\}$.
Then there exists no error pattern which is lighter than $\textbf{e}^*$.
\end{mythname}

\begin{IEEEproof} Let $\textbf{e}$ be a hypothesized error pattern. Since any two hypothesized error patterns must have Hamming distance at least $d_{\min}$, we have $\lambda(\textbf{e}) = \sum_{j \in \mathcal{S}(\textbf{e}^*)} \lambda(e_j) + \sum_{j \notin \mathcal{S}(\textbf{e}^*)} \lambda(e_j) \geq \sum_{j \notin \mathcal{S}(\textbf{e}^*)} \lambda(e_j) \geq \min_{\textbf{e}\in \mathcal{E}(\textbf{e}^*)} \lambda(\textbf{e})$.
\end{IEEEproof}

Note that the bound $B_0(\textbf{e}^*)$ can be calculated by a greedy algorithm similar to Algorithm~1.

\section{The Tree-based Chase-type GS Algorithm}\label{sec:TCGA}
From the framework of the tree-based Chase-type algorithm, we see
that the flipping pattern at the $\ell$-th trial diverges from its
parent~(which has been tested at the $i$-th trial for some $i <
\ell$) in one coordinate. This property admits a low-complexity
decoding algorithm if GS algorithm is implemented with
multiplicity one and initialized Gr\"{o}bner basis $\{1, y\}$. Let
$\deg_y(Q)$ be the $y$-degree of $Q$ and $\deg_{1,k-1}(Q)$ be the
$(1,k-1)$-weighted degree of $Q$.

\begin{myponame}\label{fpoly}
Let $\textbf{f}$ be a flipping pattern and $\mathcal{Q}(\textbf{f}) \stackrel{\Delta}{=} \{Q^{(0)}(x, y), Q^{(1)}(x, y)\}$ be the Gr\"{o}bner basis for the $\mathbb{F}_q[x]$-module $\{Q(x, y): Q(\beta_i, z_i - f_i) = 0, 0 \leq i \leq n-1, \deg_y(Q) \leq 1\}$. Then $\mathcal{Q}(\textbf{f})$ can be found from $\mathcal{Q}(\textbf{f}_\uparrow)$ by backward interpolation and forward interpolation.
\end{myponame}

\begin{IEEEproof}
Denote the two polynomials in the Gr\"{o}bner basis $\mathcal{Q}(\textbf{f}_\uparrow)$ by
$Q^{(l)}(x, y) = q_0^{(l)}(x) + q_1^{(l)}(x)y, l \in \{0,1\}$. Let the right-most atom of $\textbf{f}$ is $(j, \delta)$ which is the only atom
contained in $\textbf{f}$ but not in $\textbf{f}_\uparrow$. We need to update $\mathcal{Q}(\textbf{f}_\uparrow)$ by removing the point $(\beta_j, z_j)$ and
interpolating the point $(\beta_j, z_j - \delta)$.  This can be done according to the following steps, as shown in~\cite{zhu2009backward}.

\begin{itemize}

  \item Use backward interpolation to eliminate the point $(\beta_j, z_j)$ of $\mathcal{Q}(\textbf{f}_\uparrow)$:
\begin{enumerate}
  \item Compute $q_1^{(l)}(\beta_j)$ for $l = 0, 1$; let $\mu = \arg \min_{l} \{\deg_{1,k-1}(Q^{(l)}): q_1^{(l)}(\beta_j) \neq 0)\}$; let $\nu = 1- \mu$;
  \item If $q_1^{(\nu)}(\beta_j) \neq 0$, then $Q^{(\nu)}(x,y) \leftarrow q_1^{(\mu)}(\beta_j)Q^{(\nu)}(x,y) - q_1^{(\nu)}(\beta_j)Q^{(\mu)}(x,y)$;
      \item $Q^{(\nu)}(x,y) \leftarrow Q^{(\nu)}(x,y)/(x-\beta_j)$ and $Q^{(\mu)}(x,y) \leftarrow Q^{(\mu)}(x, y)$ ;
\end{enumerate}
  \item Use forward interpolation~(K\"{o}tter's algorithm) to add the point $(\beta_j, z_j - \delta)$:
\begin{enumerate}
  \item Compute $Q^{(l)}(\beta_j, z_j - \delta)(l = 0,1)$; let $\mu = \arg \min_l(\deg_{1, k-1}(Q^{(l)}): Q^{(l)}(\beta_j, z_j - \delta) \neq 0)$; let $\nu = 1 - \mu$;
  \item $Q^{(\nu)}(x,y) \leftarrow Q^{(\mu)}(\beta_j, z_j - \delta)Q^{(\nu)}(x,y) - Q^{(\nu)}(\beta_j, z_j - \delta)Q^{(\mu)}(x,y)$;
  \item $Q^{(\mu)}(x,y) \leftarrow Q^{(\mu)}(x,y)(x - \beta_j)$.
\end{enumerate}
\end{itemize}
To summarize, from $\mathcal{Q}(\textbf{f}_\uparrow)$, we can obtain $\mathcal{Q}(\textbf{f}) = \{Q^{(\mu)}(x,y), Q^{(\nu)}(x,y) \}$ efficiently.
\end{IEEEproof}

The main result of this section is the following tree-based Chase-type GS decoding algorithm for RS codes.

{\em \textbf{Algorithm}~3:} Tree-based Chase-type GS Decoding Algorithm for RS Codes
\begin{itemize}
  \item {\bf {Preprocessing:}} Upon on receiving the vector $\textbf{r}$, find the hard-decision vector $\textbf{z}$; compute the soft weights of all atoms; construct the atom chain $\textbf{A}$.
  \item {\bf {Initialization:}}
 \begin{enumerate}
   \item $\mathcal{F} = \textbf{0}$; $\ell = 0$; $\textbf{e}^* = \textbf{z}$; $u^*(x) = 0$;
   \item Input $\textbf{z}$ to the GS algorithm with multiplicity one and initialized Gr\"{o}bner basis $\{1, y\}$, resulting in $\mathcal{Q}(\textbf{0})$;
   \item Factorize the minimal polynomial in $\mathcal{Q}(\textbf{0})$. If a message polynomial $u(x)$ is found, find $\textbf{e} = \textbf{z} - \textbf{c}$, where $\textbf{c}$ is the codeword corresponding to $u(x)$; if $\lambda(\textbf{e}) < \lambda(\textbf{e}^*)$, set $\textbf{e}^* = \textbf{e}$ and $u^*(x) = u(x)$; if, additionally, $\lambda(\textbf{e}) \leq B_0(\textbf{e})$, output $u(x)$ and exit the algorithm;
    \item Insert $\textbf{0}_\downarrow$~(the left-most child of $\textbf{f}$) into $\mathcal{F}$ and remove $\textbf{0}$ from $\mathcal{F}$; set $\mathcal{Q}(\textbf{0}_\downarrow) = \mathcal{Q}(\textbf{0})$;
   \item Set $\ell = 1$.
 \end{enumerate}
  \item {\bf{Iterations:}} While $\ell < L$, do the following.
 \begin{enumerate}
   \item Set $\textbf{f} = \textbf{f}^{(\ell)}$; if $\lambda(\textbf{e}^*) \leq B(\textbf{f})$, output $u^*(x)$ and exit the algorithm;
   \item Let $(j, \delta)$ be the right-most atom of $\textbf{f}$. Update $\mathcal{Q}(\textbf{f})$ by removing the point $(\beta_j, z_j)$ and interpolating the point $(\beta_j, \delta)$;
   \item Factorize the minimal polynomial in $\mathcal{Q}(\textbf{f})$. If a message polynomial $u(x)$ is found, find $\textbf{e} = \textbf{z} - \textbf{c}$, where $\textbf{c}$ is the codeword corresponding to $u(x)$; if $\lambda(\textbf{e}) < \lambda(\textbf{e}^*)$, set $\textbf{e}^* = \textbf{e}$ and $u^*(x) = u(x)$; if, additionally, $\lambda(\textbf{e}) \leq B_0(\textbf{e})$, output $u(x)$ and exit the algorithm;
   \item Insert~(if exist) $\textbf{f}_\downarrow$~(the left-most child of $\textbf{f}$) and  $\textbf{f}_\rightarrow$~(the following right-sibling of $\textbf{f}$) into $\mathcal{F}$; set $\mathcal{Q}(\textbf{f}_\downarrow) = \mathcal{Q}(\textbf{f})$ and $\mathcal{Q}(\textbf{f}_\rightarrow) = \mathcal{Q}(\textbf{f}_\uparrow)$; remove $\textbf{f}^{(\ell)}$ from the linked list $\mathcal{F}$;
   \item Increment $\ell$ by one.
 \end{enumerate}
\end{itemize}

{\bf Remark.}
\begin{itemize}
\item It is worth pointing out that the factorization step in Algorithm~3 can be implemented in a simple way as shown in~\cite{McEliece03}. Let $q_0(x) + q_1(x)y$ be the polynomial to be
factorized. If $q_1(x)$ divides $q_0(x)$, set $u(x) = - q_{0}(x) / q_1(x)$. If ${\rm deg} (u(x)) < k$, $u(x)$ is a valid message polynomial.

\item Note that, to use the backward interpolation techniques in Algorithm 3, besides maintaining the linked list $\mathcal{F}$, we have to reserve these interpolation polynomials corresponding to their parents. Since the size of $\mathcal{F}$ can be kept as less than or equal to $L-\ell$ by removing extra tailed flipping patterns, maintaining the linked list $\mathcal{F}$ requires extra reservation at most $O((L-\ell)n)$. By observing the step~2) and~3) in backward and forward interpolation, we can conclude that its complexity is roughly $O(n)$ at each iteration.
\end{itemize}

To illustrate clearly the construction of the tree $\textbf{T}$ as well as the tree-based Chase-type GS decoding algorithm, we give below an example.

\begin{myexname}
Consider the RS code $\mathcal{C}_5[4, 2]$ over $\mathbb{F}_5 = \{0,1 ,2 ,3, 4\}$ with $t_{\min} = 1$. Let the message polynomial $u(x)\! =\! 1 + 2x$ and $\mathcal{L}\! =\! \{0,1,2,3\}$.
Then the codeword $\textbf{c}\! =\! (u(0), u(1), u(2), u(3)) = (1, 3, 0, 2)$.
Let $\textbf{r}$ be the received vector from a memoryless channel that specifies the following log-likelihood matrix
$$\Pi \!=\! [\pi_{i,j}]\! =\! \left[\renewcommand{\arraystretch}{0.2}\begin{array}{cccc}
-2.44 &  -1.41 &  -1.37 &  -1.45 \\
-1.20 &  -1.87 &  -3.24 &  -2.18 \\
-2.76 &  -1.50 &  -1.22 &  -1.56 \\
-2.32 &  -1.63 &  -2.64 &  -1.48 \\
-1.45 &  -2.35 &  -1.81 &  -1.77
\end{array}\right]\!,$$
where $\pi_{i, j} = \log {\rm Pr}(r_j| i)$ for $0 \leq i \leq 4$ and $0 \leq j \leq 3$.

The tree-based Chase-type GS decoding algorithm with $L = 16$ is performed as follows.

{\bf{\em Preprocessing:}} Given the log-likelihood matrix $\Pi$, find the hard-decision vector $\textbf{z} = (1, 0, 2, 0)$. Find the soft weights of all atoms, which can be arranged as
$$\Lambda = [\lambda_{i, j}] = \left[\renewcommand{\arraystretch}{0.2}\begin{array}{cccc}
1.24 &  0.94 &  2.02 &  0.32 \\
0.25 &  0.22 &  0.15 &  0.03 \\
1.12 &  0.09 &  0.59 &  0.11 \\
1.56 &  0.46 &  1.42 &  0.73
\end{array}\right],$$
where $\lambda_{i, j} = \lambda_j(i) = \log {\rm Pr}(r_j|z_j) - \log {\rm Pr}(r_j|z_j - i)$ is the soft weight of atom $(j, i)$ for $1 \leq i \leq 4$ and $0 \leq j \leq 3$. Given $\Lambda$, all the $16$ atoms
can be arranged into the atom chain $\textbf{A} = [(3,2) \prec (1,3) \prec (3,3) \prec (2,2) \prec (1,2) \prec (0,2) \prec (3,1) \prec (1,4) \prec (2,3) \prec (3,4) \prec (1,1) \prec
(0,3) \prec (0,1) \prec (2,4) \prec (0,4) \prec (2,1)].$

{\bf {\em Initialization:}}
 \begin{enumerate}
   \item $\mathcal{F} = \textbf{0}$; $\ell = 0$; $\textbf{e}^* = \textbf{z} = (1, 0, 2, 0)$; $\lambda(\textbf{e}^*) = 1.39$; $u^*(x) = 0$;
   \item input $\textbf{z} = (1, 0, 2, 0)$ to the GS decoder and obtain $\mathcal{Q}(\textbf{0}) = \{4 + 2x + x^2 + 3x^3 + (1 + 3x)y, 1 + 2x + 2x^2 + (4 + x)y\}$;
   \item factorize $1 + 2x + 2x^2 + (4 + x)y$; since $1 + 2x + 2x^2$ is divisible by $4 + x$, find a valid message polynomial $u(x) = 1 + 3x$, which generates the codeword $\textbf{c} = (1, 4, 2, 0)$; obtain $\textbf{e} = \textbf{z} - \textbf{c} = (0, 1, 0, 0)$ and $\lambda(\textbf{e}) = 0.94$; since $\lambda(\textbf{e}) < \lambda(\textbf{e}^*)$, we set $\textbf{e}^* = (0, 1, 0, 0)$ and $u^*(x) = u(x)$; since $\lambda(\textbf{e}) > B_0(\textbf{e})~ (= 0.18)$, the algorithm continues;
   \item insert $(3,2)$~(the left-most child of $\textbf{0}$) into $\mathcal{F}$ as shown in Fig.~\ref{CT01}-(1) and remove $\textbf{0}$ from $\mathcal{F}$; set $\mathcal{Q}(\textbf{0}_\downarrow) = \mathcal{Q}(\textbf{0})$;
       at this step, the linked list $\mathcal{F} = \underbrace{\textbf{0}
   \prec}_{\rm removed}
   \textbf{f}^{(1)}$ with $\textbf{f}^{(1)} = (3,2)$;
   \item set $\ell = 1$;
 \end{enumerate}

 \begin{figure}
\centering
\includegraphics[width=8.0cm]{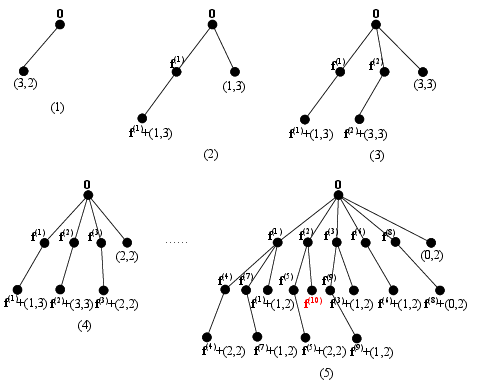}
\begin{center}
\caption{An example of constructing progressively the tree of flipping patterns.}\label{CT01}
\end{center}
\end{figure}

{\bf{\em Iterations:}} While $\ell < 16$, do\\
When $\ell = 1$,
 \begin{enumerate}
   \item set $\textbf{f} = \textbf{f}^{(1)} = (3, 2)$;  since $\lambda(\textbf{e}^*) = 0.94 > B(\textbf{f}) = 0.12$, the algorithm continues;
   \item since the right-most atom of $\textbf{f}$ is $(3, 2)$, update $\mathcal{Q}(\textbf{f})$ by removing $(3, 0)$ and interpolating $(3, 3)$ and obtain $\mathcal{Q}(\textbf{f}) = \{4 + 2x + x^2 + 3x^3 + (1  + 3x)y , 4 + 4x + 2x^2 + (1 + 2x)y\}$;
   \item factorize $4 + 4x + 2x^2 + (1 + 2x)y$; since $4 + 4x + 2x^2$ is divisible by $1 + 2x$, find a valid message polynomial $u(x) = 1 + 4x$, which generates the codeword $\textbf{c} = (1, 0, 4, 3)$; obtain $\textbf{e} = \textbf{z} - \textbf{c} = (0, 0, 3, 2)$ and compute $\lambda(\textbf{e}) = 0.62$; since $\lambda(\textbf{e}) < \lambda(\textbf{e}^*) (= 0.94)$, we set $\textbf{e}^* = (0, 0, 3, 2)$ and $u^*(x) = u(x)$; since $\lambda(\textbf{e}) > B_0(\textbf{e})~(= 0.09)$, the algorithm continues;
   \item insert $\textbf{f}_\downarrow = \textbf{f} + (1, 3)$ and  $\textbf{f}_\rightarrow = (1, 3)$ into $\mathcal{F}$ as shown in Fig.~\ref{CT01}-(2); set $\mathcal{Q}(\textbf{f}_\downarrow) = \mathcal{Q}(\textbf{f})$ and $\mathcal{Q}(\textbf{f}_\rightarrow) = \mathcal{Q}(\textbf{0})$; remove $\textbf{f}^{(1)}$ from the linked list $\mathcal{F}$;
       at this step, the linked list $\mathcal{F} = \underbrace{\textbf{0}
   \prec \textbf{f}^{(1)} \prec}_{\rm removed}~\textbf{f}^{(2)} \prec \textbf{f}^{(1)} + (1,3)$, where $\textbf{f}^{(2)} = (1, 3)$;
   \item set $\ell = 2$.
 \end{enumerate}
When $\ell = 2$,
 \begin{enumerate}
   \item set $\textbf{f} = \textbf{f}^{(2)} = (1, 3)$; since $\lambda(\textbf{e}^*) = 0.62 > B(\textbf{f}) = 0.20$, the algorithm continues;
   \item since the right-most atom of $\textbf{f}$ is $(1, 3)$, update $\mathcal{Q}(\textbf{f})$ by removing $(1, 0)$ and interpolating $(1, 2)$ and obtain $\mathcal{Q}(\textbf{f}) = \{3x^2 + 4x^3 + 4xy , 1 + 2x + 2x^2 + (4 + x)y\}$;
   \item factorize $1 + 2x + 2x^2 + (4 + x)y$; since $1 + 2x + 2x^2$ is divisible by $4 + x$, find a valid message polynomial $u(x) = 1 + 3x$,
   which generates the codeword $\textbf{c} = (1, 4, 2, 0)$; obtain $\textbf{e} = \textbf{z} - \textbf{c} = (0, 1, 0, 0)$ and
   $\lambda(\textbf{e}) = 0.94$; since $\lambda(\textbf{e}) \geq \lambda(\textbf{e}^*)~(= 0.62)$, updating $\textbf{e}^*$ and $u^*(x)$
   are not required;
   \item insert $\textbf{f}_\downarrow = \textbf{f} + (3, 3)$ and  $\textbf{f}_\rightarrow = (3, 3)$ into $\mathcal{F}$ as shown in Fig.~\ref{CT01}-(3); set $\mathcal{Q}(\textbf{f}_\downarrow) = \mathcal{Q}(\textbf{f})$ and $\mathcal{Q}(\textbf{f}_\rightarrow) = \mathcal{Q}(\textbf{0})$; remove $\textbf{f}^{(2)}$ from the linked list $\mathcal{F}$;
       at this step, the linked list $\mathcal{F} = \underbrace{\textbf{0}
   \prec \textbf{f}^{(1)} \prec \textbf{f}^{(2)} \prec }_{\rm removed}~\textbf{f}^{(3)} \prec \textbf{f}^{(1)} + (1,3) \prec \textbf{f}^{(2)} + (3,3)$, where $\textbf{f}^{(3)} = (3, 3)$;
   \item set $\ell = 3$.
 \end{enumerate}
When $\ell = 3$,
 \begin{enumerate}
   \item set $\textbf{f} = \textbf{f}^{(3)} = (3, 3)$; since $\lambda(\textbf{e}^*) = 0.62 > B(\textbf{f}) = 0.26$, the algorithm continues;
   \item since the right-most atom of $\textbf{f}$ is $(3, 3)$, update $\mathcal{Q}(\textbf{f})$ by removing $(3, 0)$ and interpolating $(3, 2)$ and obtain $\mathcal{Q}(\textbf{f}) = \{2 + 3x^2 + 3y, 2 + x^2 + 2x^3 + (3 + x + x^2)y\}$;
   \item factorize $2 + 3x^2 + 3y$; no candidate codeword is found at this step;
   \item insert $\textbf{f}_\downarrow = \textbf{f} + (2, 2)$ and  $\textbf{f}_\rightarrow = (2, 2)$ into $\mathcal{F}$ as shown in Fig.~\ref{CT01}-(4); set $\mathcal{Q}(\textbf{f}_\downarrow) = \mathcal{Q}(\textbf{f})$ and $\mathcal{Q}(\textbf{f}_\rightarrow) = \mathcal{Q}(\textbf{0})$; remove $\textbf{f}^{(3)}$ from the linked list $\mathcal{F}$;
       at this step, the linked list $\mathcal{F} = \underbrace{\textbf{0}
   \prec \textbf{f}^{(1)} \prec \textbf{f}^{(2)} \prec \textbf{f}^{(3)} \prec }_{\rm removed}~\textbf{f}^{(4)} \prec \textbf{f}^{(2)} + (3,3) \prec (2, 2) \prec \textbf{f}^{(3)} + (2, 2)$, where $\textbf{f}^{(4)} = \textbf{f}^{(1)} + (1, 3) = (3, 2) + (1, 3)$;
   \item set $\ell = 4$.
 \end{enumerate}
$$~~~~\vdots~~~~~~~~~~~~~~~~~~~~~~~~~~~~~~~~~~~~~~~~~~~~~~~~~~~~~~~~~~~~~~~~~~~~~~~~~~~~~~~~~~~~~~~~~~~~~~~~~~$$
When $\ell = 9$,
 \begin{enumerate}
    \item set $\textbf{f} = \textbf{f}^{(9)} = (3, 3) + (2, 2)$; since $\lambda(\textbf{e}^*) = 0.62 < B(\textbf{f}) = 0.48$, the algorithm continues;
    \item since the right-most atom of $\textbf{f}$ is $(2, 2)$, update $\mathcal{Q}(\textbf{f})$ by removing $(2, 2)$ and interpolating $(2, 0)$ and obtain $\mathcal{Q}(\textbf{f}) = \{3 + 2x + 3x^2 + 2x^3 + (2 + x)y , 2 + 2x + x^2 + (3 + 2x)y\}$;
    \item factorize $2 + 2x + x^2 + (3 + 2x)y$; since $2 + 2x + x^2$ is divisible by $3 + 2x$, find a valid message polynomial $u(x) = 1 + 2x$, which generates the codeword $\textbf{c} = (1, 3, 0, 2)$; obtain $\textbf{e} = \textbf{z} - \textbf{c} = (0, 2, 2, 3)$ and $\lambda(\textbf{e}) = 0.48$; since $\lambda(\textbf{e}) < \lambda(\textbf{e}^*) (= 0.62)$, set $\textbf{e}^* = (0, 2, 2, 3)$ and $u^*(x) = u(x)$; since $\lambda(\textbf{e}) > B_0(\textbf{e})~(= 0)$, the algorithm continues;
    \item insert $\textbf{f}_\downarrow = \textbf{f} + (1, 2)$ and  $\textbf{f}_\rightarrow = (3, 3) + (1, 2)$ into $\mathcal{F}$ as shown in Fig.~\ref{CT01}-(5); set $\mathcal{Q}(\textbf{f}_\downarrow) = \mathcal{Q}(\textbf{f})$ and $\mathcal{Q}(\textbf{f}_\rightarrow) = \mathcal{Q}(\textbf{f}_\uparrow)$ with $\textbf{f}_\uparrow = (3,3)$; remove $\textbf{f}^{(9)}$ from the linked list $\mathcal{F}$;
        at this step, the linked list $\mathcal{F} = \underbrace{\textbf{0}
   \prec \textbf{f}^{(1)} \prec \cdots \prec \textbf{f}^{(9)} \prec}_{\rm removed}~\textbf{f}^{(10)} \prec \textbf{f}^{(1)} + (1, 2) \prec \cdots \prec \textbf{f}^{(8)} + (0, 2)$, where $\textbf{f}^{(10)} = \textbf{f}^{(2)} + (2, 2) = (1, 3) + (2, 2)$;
   \item update $\ell = 10$.
 \end{enumerate}
 When $\ell = 10$,
 \begin{enumerate}
    \item set $\textbf{f} = \textbf{f}^{(10)} = (1, 3) + (2, 2)$; since $\lambda(\textbf{e}^*) = 0.48 < B(\textbf{f}) = 0.49$, output $u^*(x) = 1 + 2x$ and exit the algorithm.
 \end{enumerate}
\hfill{$\square$}
\end{myexname}

\section{Numerical Results and Further Discussions}\label{sec:PACC}

In this section, we compare the proposed tree-based Chase-type GS~(TCGS)
decoding algorithm with the LCC decoding algorithm~\cite{Bellorado10}. We take the LCC decoding algorithm as a benchmark since the TCGS algorithm is similar to the LCC algorithm with the exception of the set of flipping patterns and the testing orders. In all examples, messages are encoded by RS codes and then transmitted over additive white Gaussian noise~(AWGN) channels with binary phase shift keying~(BPSK) modulation. The performance is measured by the frame error rate~(FER), while the complexity is measured in terms of the average testing numbers. For a fair and reasonable comparison, we assume that these two algorithms perform the same maximum number of trials, that is, $L = 2^{\eta}$. The LCC decoding algorithm takes Theorem~\ref{Kanekolemma} as the early stopping criterion, while the TCGS decoding algorithm takes both Theorem~\ref{removebound} and Theorem~\ref{Kanekolemma} as the early stopping criteria. Notice that Theorem~\ref{removebound} is one inherent feature of the TCGS algorithm, which does not apply to the LCC algorithm. For reference, the performance of the GMD decoding algorithm and that of the theoretical KV decoding algorithm~(with an infinite interpolation multiplicity) are also given.

\subsection{Numerical Results}

\begin{figure}
\centering
\includegraphics[width=8.0cm]{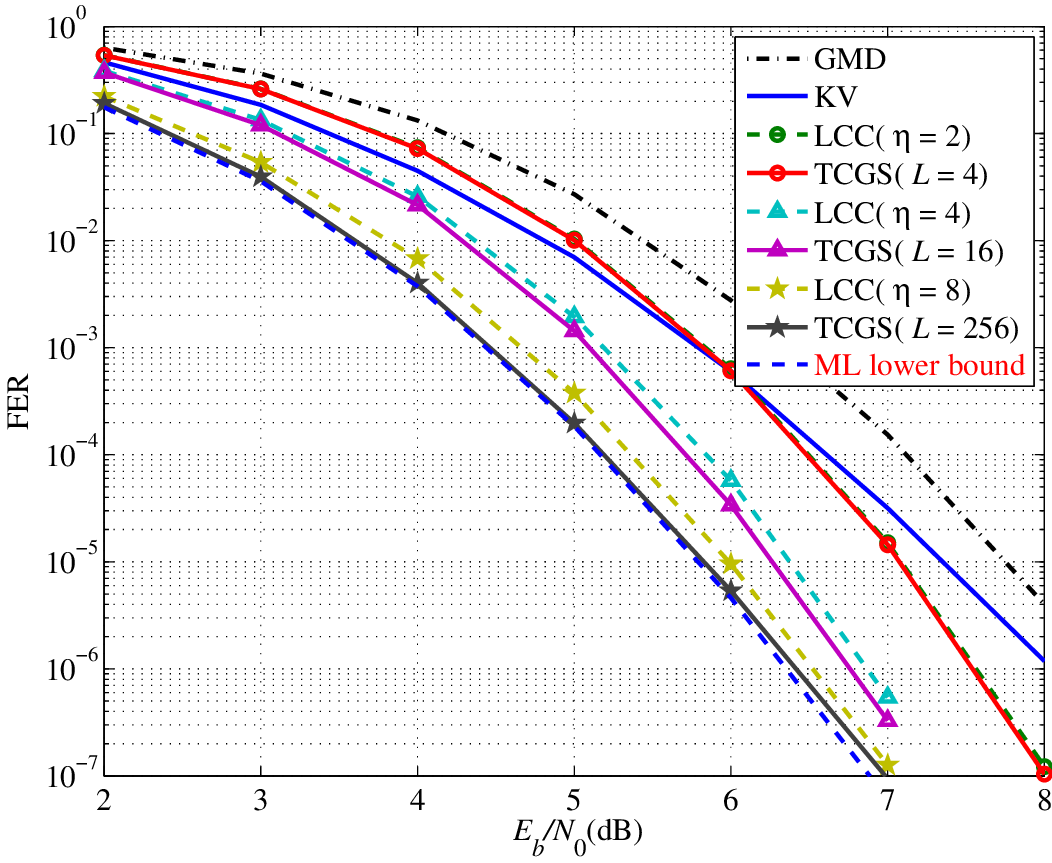}
\begin{center}
\caption{Performance of the tree-based Chase-type decoding of the RS code $\mathcal{C}_{16}[15, 11]$.}\label{CT1511fer}
\end{center}
\end{figure}

\begin{figure}
\centering
\includegraphics[width=8.0cm]{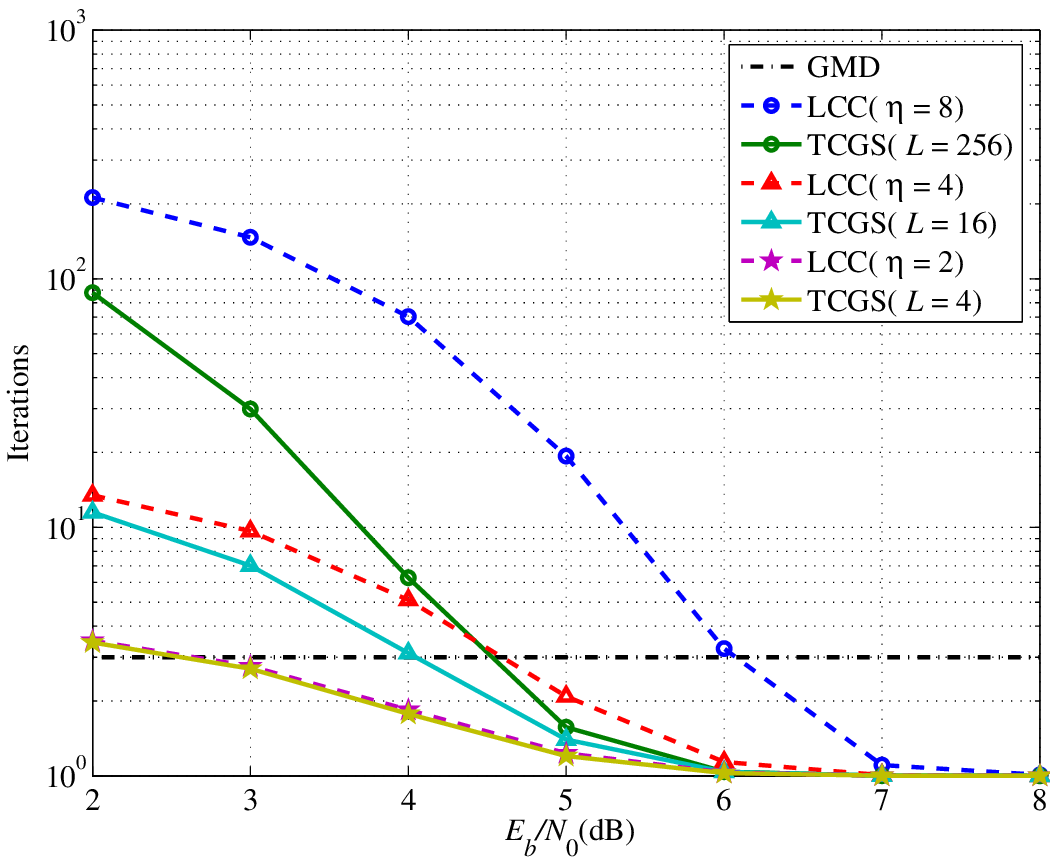}
\begin{center}\vspace{-0.05cm}
\caption{Complexity of the tree-based Chase-type decoding of the RS code $\mathcal{C}_{16}[15, 11]$.}\label{CT1511c}
\end{center}
\end{figure}

\begin{myexname}
Consider the RS code $\mathcal{C}_{16}[15, 11]$ over $\mathbb{F}_{16}$ with $t_{\min} = 2$. The performance curves are shown in Fig.~\ref{CT1511fer}. We can see that the TCGS algorithm performs slightly better than the LCC algorithm. As $L = 2^\eta$ increases, the gap becomes larger. At FER = $10^{-5}$, the TCGS algorithm with $L = 256$ outperforms the LCC algorithm~(with $\eta = 8$) and the GMD algorithm by $0.2$~dB and $2.0$~dB, respectively. Also note that, even with small number of trials, the TCGS algorithm can be superior to the KV algorithm.

The average iterations are shown in Fig.~\ref{CT1511c}. It can be seen that the average decoding complexity of both the TCGS and the LCC algorithms decreases as the SNR increases. The TCGS
algorithm requires less average iterations than the LCC algorithm. Furthermore, the average iterations
required for the TCGS algorithm are even less than those for the GMD algorithm when ${\rm SNR} \geq 5.0$~dB.
\hfill{$\square$}
\end{myexname}

\begin{figure}
\centering
\includegraphics[width=8.0cm]{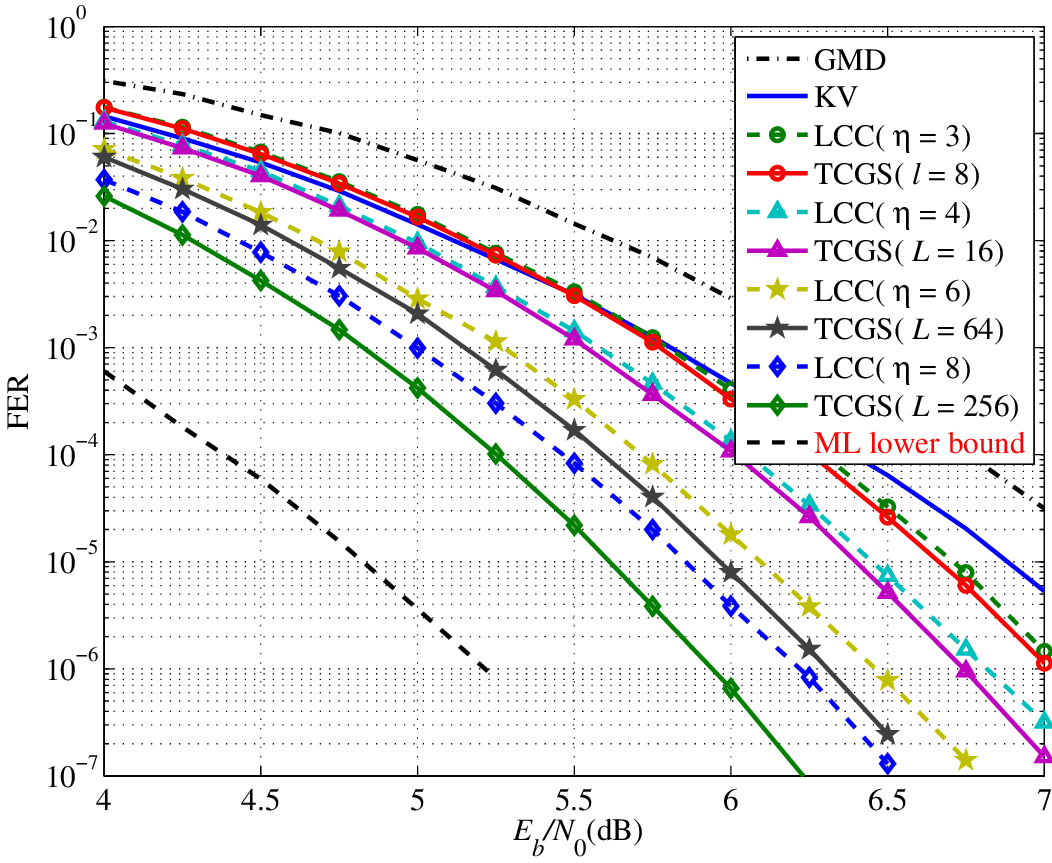}
\begin{center}
\caption{Performance of the tree-based Chase-type decoding of the RS code $\mathcal{C}_{32}[31, 25]$.}\label{CT3125fer}
\end{center}
\end{figure}

\begin{figure}
\centering
\includegraphics[width=8.0cm]{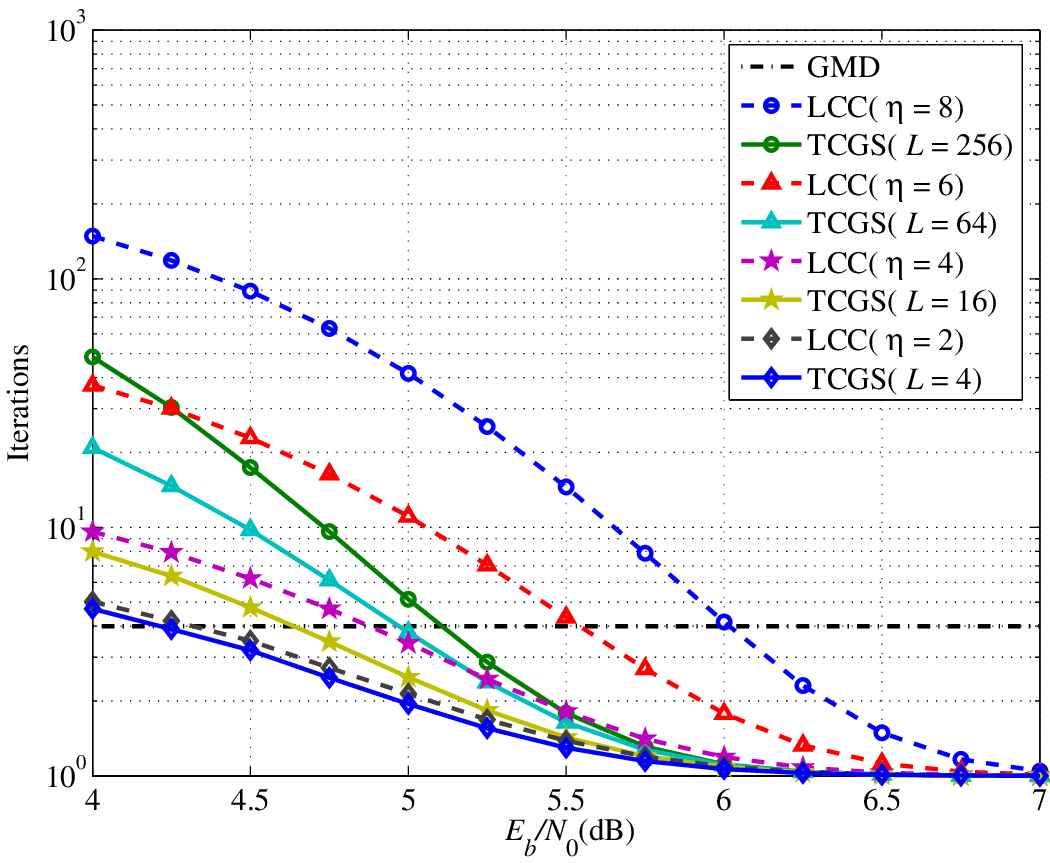}
\begin{center}\vspace{-0.05cm}
\caption{Complexity of the tree-based Chase-type decoding of the RS code $\mathcal{C}_{32}[31, 25]$.}\label{CT3125c}
\end{center}
\end{figure}

\begin{myexname}
Consider the RS code $\mathcal{C}_{32}[31, 25]$ over $\mathbb{F}_{32}$ with $t_{\min} = 3$. The performance curves are shown in Fig.~\ref{CT3125fer}. We can see that the TCGS algorithm performs slightly better than the LCC algorithm. As $L = 2^\eta$ increases, the gap becomes larger. At FER = $10^{-5}$, the TCGS algorithm with $L = 256$ outperforms the LCC algorithm~(with $\eta = 8$) and the GMD algorithm by $0.25$~dB and $1.5$~dB, respectively. Also note that, even with small number of trials, the TCGS algorithm can be superior to the KV algorithm.

The average iterations are shown in Fig.~\ref{CT3125c}. It can be seen that the average decoding complexity of both the TCGS and the LCC algorithms decreases as the SNR increases. The TCGS algorithm requires less average iterations than the LCC algorithm. Furthermore, the average iterations required for the TCGS algorithm are even less than those for the GMD algorithm when ${\rm SNR} \geq 5.25$~dB.
\hfill{$\square$}
\end{myexname}

\begin{figure}
\centering
\includegraphics[width=8.0cm]{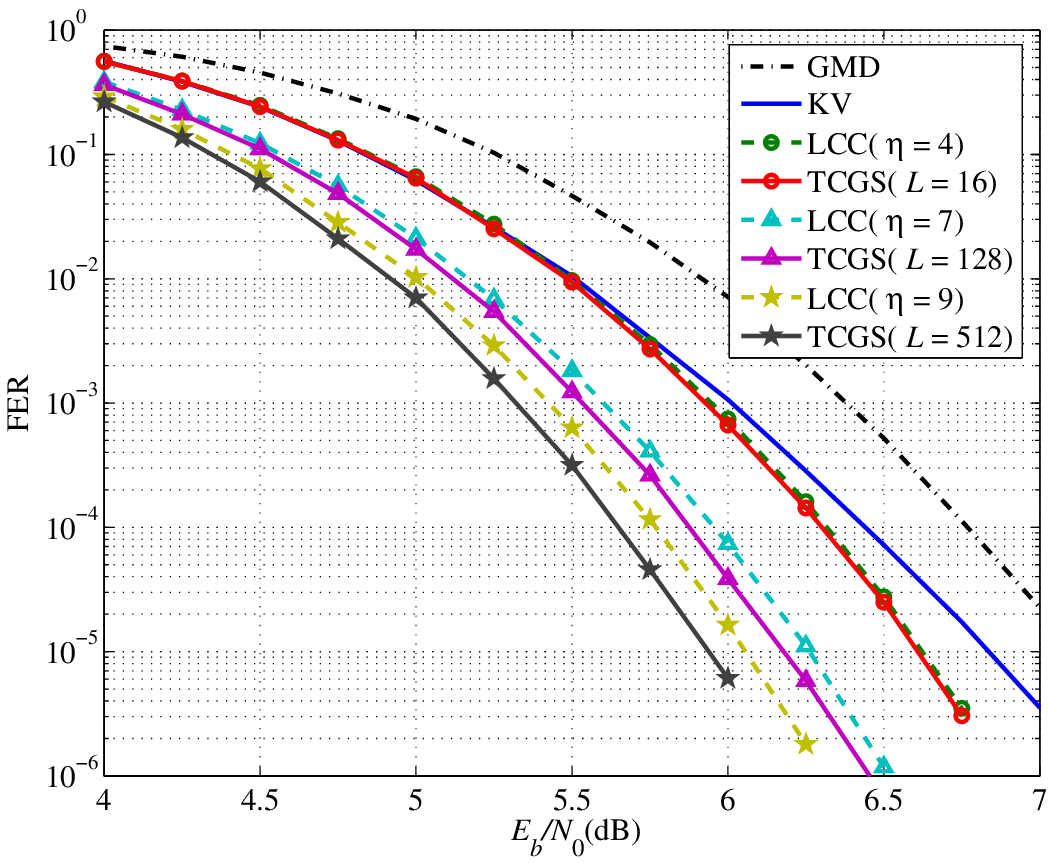}
\begin{center}
\caption{Performance of the tree-based Chase-type decoding of the RS code $\mathcal{C}_{64}[63, 55]$.}\label{CT6355fer}
\end{center}
\end{figure}

\begin{figure}
\centering
\vspace{0.06cm}
\includegraphics[width=8.0cm]{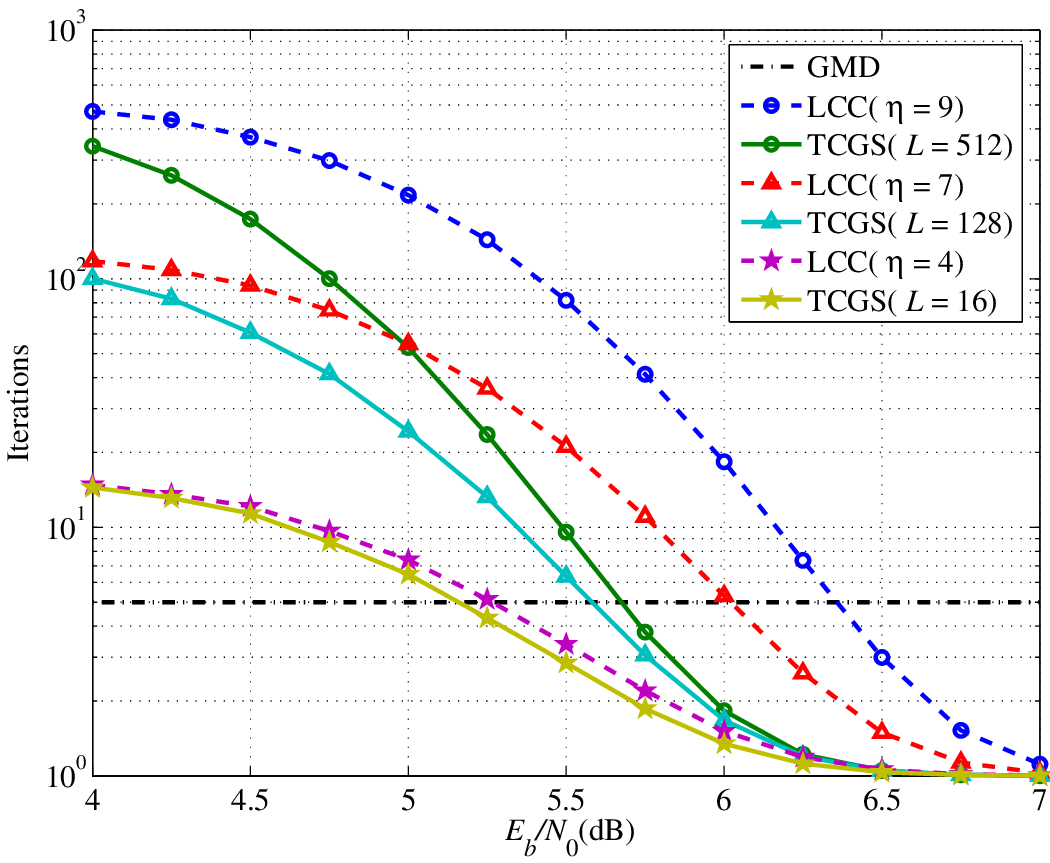}\hspace{5cm}
\begin{center}
\vspace{-0.05cm}
\caption{Complexity of the tree-based Chase-type decoding of the RS code $\mathcal{C}_{64}[63, 55]$.}\label{CT6355c}
\end{center}
\end{figure}

\begin{myexname}
Consider the RS code $\mathcal{C}_{64}[63, 55]$ over $\mathbb{F}_{64}$ with $t_{\min} = 4$. The performance curves are shown in Fig.~\ref{CT6355fer}. We can see that the TCGS algorithm performs slightly better than the LCC algorithm. As $L = 2^\eta$ increases, the gap becomes larger. At FER = $10^{-5}$, the TCGS algorithm with $L = 512$ outperforms the LCC algorithm~(with $\eta = 9$) and the GMD algorithm by $0.2$~dB and $1.2$~dB, respectively. Also note that, even with small number of trials, the TCGS algorithm can be superior to the KV algorithm.

The average iterations are shown in Fig.~\ref{CT6355c}. It can be seen that the average decoding complexity of both the TCGS and the LCC algorithms decreases as the SNR increases. The TCGS algorithm requires less average iterations than the LCC algorithm. Furthermore, the average iterations required for the TCGS algorithm are even less than those for the GMD algorithm when ${\rm SNR} \geq 5.75$~dB.
\hfill{$\square$}
\end{myexname}

\subsection{Complexity Analysis}
In summary, we have compared by simulation the tree-based Chase-type GS (TCGS) decoding algorithm with the LCC decoding algorithm, showing that the TCGS algorithm has a better performance and requires less trials for a given maximum testing number. We will not argue that the proposed algorithm has lower complexity since it is difficult to make such a comparison. The difficulty lies in that, although the interpolation process in the finite field is simple, the proposed algorithm requires pre-processing and evaluating the lower bound for each flipping pattern in the real field. As mentioned above, the computation of preprocessing step is $O(n(q-1)\log(n(q-1)))$, and to maintain the linked list $\mathcal{F}$ requires computational complexity of order at most $O(L\log L)$. For each iteration, compared to the ordered rooted tree generation, the complexity is dominated by the HDD of $L$ testing patterns.

Although a detailed analysis of the computational complexity of the total decoding process is difficult, the upper bound on the number of finite field arithmetic operations (i.e., additions and multiplications) for decoding one codeword frame can be provide to measure the decoding complexity. According to literature~\cite{tang2013progressive}~\cite{Bellorado10}, complexity bounds for interpolation and factorization are given as follows:
\begin{equation}\label{cint}
\mathbb{C}_{int} \leq 2n^2 + 2n(L-1)
\end{equation}
\begin{equation}\label{cfac}
\mathbb{C}_{fac} \leq knL
\end{equation}
By replacing $L$ with $2^{\eta}$, the above complexity bounds are also the maximum computational cost of the LCC algorithm. By summing the maximum number of finite field arithmetic operations required for each step of GMD~\cite{Kotter96}, we obtain the following bound:
\begin{equation}\label{cGMD}
\mathbb{C}_{GMD} \leq (n^2 + (n-k)(7n-4k) + n)(\lfloor (n-k)/2\rfloor + 1)
\end{equation}

In TABLE~\ref{t1}, we present a complexity comparison of GMD and TCGS decoding for the following codes: $\mathcal{C}_{16}[15, 11], \mathcal{C}_{32}[31, 25], \mathcal{C}_{64}[63, 55]$. We give results of $L = 4,16$, because these values of $L$ provide a significant performance gain over GMD without a large increase in complexity. As can be seen, TCGS or LCC decoding even perform better than GMD with less complexity when $L \leq 4$ or $\eta \leq 2$.

\begin{table}[h]
  \centering
  \caption{UPPER BOUND ON THE NUMBER OF FINITE FIELD ARITHMETIC OPERATIONS TO DECODE A SINGLE FRAME FOR GMD AND TCGS DECODING}\label{t1}
\small
\begin{tabular}{|c|c|c|c|}
\cline{1-4}
  ALGORITHM & $\mathcal{C}_{16}[15, 11]$ & $\mathcal{C}_{32}[31, 25]$ & $\mathcal{C}_{64}[63, 55]$ \\
\cline{1-4}
  GMD & 1452 & 6664 & 29000 \\
\cline{1-4}
  TCGS($L$ = 4) & 1200 & 5208 & 22176 \\
\cline{1-4}
  TCGS($L$ = 16) & 3540 & 15252 & 65268 \\
\cline{1-4}
\end{tabular}
\end{table}

Moreover, the proposed algorithm have the figure of merits as discussed in the following subsections.

\subsection{Further Discussions}
\begin{figure}
\centering
\includegraphics[width=8.0cm]{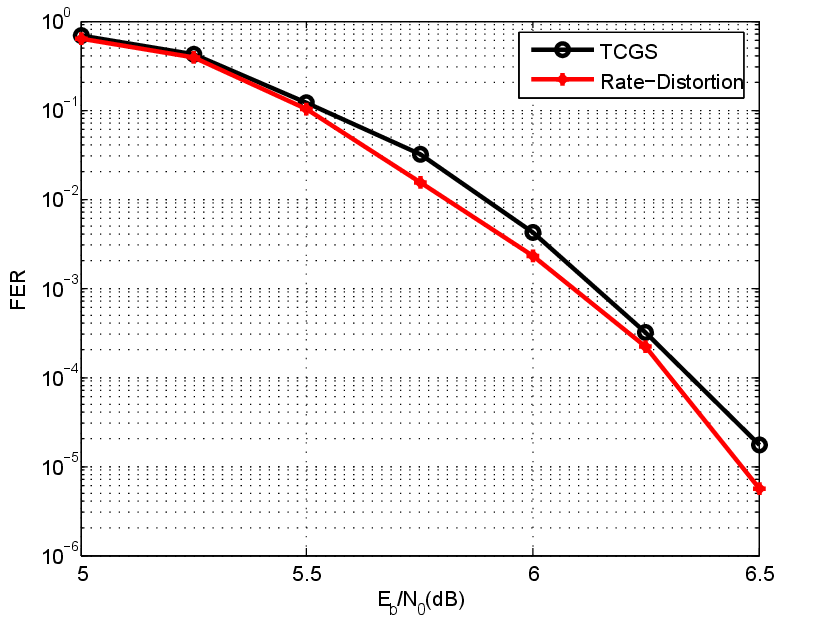}
\begin{center}
\caption{Performance of the tree-based Chase-type decoding of the RS code $\mathcal{C}_{256}[255, 239]$.}\label{CT255239}
\end{center}
\end{figure}

To compare with the rate-distortion approach used in~\cite{Nguyen11}, in Fig.~\ref{CT255239}, we also consider 256ASD-3(RDE,11) algorithm of~\cite{Nguyen11}, in which the ASD decoding is performed  $2^{11}$ times, and for each time the computational complexity of the ASD algorithm is roughly $O(n^2)$~\cite{McEliece03}. When $\eta = 11$, the TCGS algorithm has the same maximum number of decoding trials as 256ASD-3(RDE,11) algorithm. Although the former has a little worse performance than the latter, the former requires both less decoding complexity per trial and less average iterations than the latter.

Most existed Chase-type algorithms set a combinatorial number as the maximum testing number. In contrast, the proposed algorithm can take any positive integer as the maximum testing number. Even better, we can set a threshold of the soft weight to terminate the algorithm, that searches the most-likely codeword within an Euclidean sphere rather than a Hamming sphere. The larger the soft discrepancy threshold, the smaller the performance loss. This TCGS algorithm presents a method to simulate the ML lower bound, which is very similar to the one introduced in~\cite{Chase72}. If the transmitted codeword is less likely than the estimated codeword from the TCGS decoder, then the ML decoder will surely make an error in this instance as well. We count such instances to generate the lower bound. The ML lower bounds are shown in both Fig.~\ref{CT1511fer} and Fig.~\ref{CT3125fer}. In Fig.~\ref{CT1511fer}, we see that, for $\mathcal{C}_{16}[15, 11]$, the performance of TCGS with $L = 256$ is very close to the ML lower bound. While for $\mathcal{C}_{32}[31, 25]$, the performance of TCGS with $L = 256$ is about 0.7~dB away from the ML lower bound in Fig.~\ref{CT3125fer}. In the simulation, we observe that, frequently, the maximum likelihood candidate occurs much earlier than it is identified. This suggests us to improve the sufficient condition further, as stimulates our future work.

\section{Conclusions}\label{sec:Conclusions}
In this paper, we have presented a new Chase-type soft-decision decoding algorithm for RS codes. The
key to develop such an algorithm is the arrangement of all possible flipping patterns in an ordered  rooted tree according to the associated lower bounds on their soft weights. With this tree, all flipping patterns can be sorted in a serial manner, which reduces the complexity efficiently when combining with the presented sufficient conditions for optimality. Simulation results show that the proposed tree-based Chase-type algorithm performs well even with less trials on average.

\section*{Acknowledgment}
The authors are grateful to Mr. Yuan Zhu and Dr. Li Chen for useful discussions.

\appendices
\section{The Minimality of the Minimal Decomposition}\label{sec:Minimalityoffp}
There are exactly $q^k$ hypothesized error patterns, which collectively form a coset of the code, $\mathcal{E} = \{\textbf{e} = \textbf{z} - \textbf{c}: \textbf{c} \in \mathcal{C}\}$. For every $\textbf{f} \in \mathbb{F}_q^n$, when taking $\textbf{z} - \textbf{f}$ as an input vector, the conventional HDD either reports a decoding failure or outputs a unique codeword $\textbf{c}$. In the latter case, we say that the flipping pattern $\textbf{f}$ {\em generates} the hypothesized error pattern $\textbf{e} = \textbf{z} - \textbf{c}$. It can be verified that there exist $\sum_{0\leq t \leq t_{\min} }\binom{n}{t} q^{t}$ flipping patterns that can generate the same hypothesized error pattern. In fact, for each $\textbf{e} \in \mathcal{E}$, define $\mathcal{F}(\textbf{e}) = \{\textbf{f}: \textbf{e} = \textbf{f} + \textbf{g}, W_H(\textbf{g})\leq t_{\min}\}$, which consists of all flipping patterns that generate $\textbf{e}$.

\begin{myponame}\label{ps02}
Let $\textbf{e}$  be a hypothesized error pattern and $\textbf{f}$ be its associated minimal flipping pattern. Then $\textbf{f} \in \mathcal{F}(\textbf{e})$, $\lambda(\textbf{f}) = \min_{\textbf{h}\in \mathcal{F}(\textbf{e})} \lambda(\textbf{h})$ and $R_u(\textbf{f}) = \min_{\textbf{h}\in \mathcal{F}(\textbf{e})} R_u(\textbf{h})$.
\end{myponame}
\begin{IEEEproof}
Since $\textbf{e}$ is a hypothesized error pattern, $\textbf{c} = \textbf{z} - \textbf{e}$ is a codeword, which can definitely be found whenever the HDD takes as input a vector $\textbf{c} + \textbf{g}$ with $W_H(\textbf{g})\leq t_{\min}$. That is, $\textbf{e}$ can be generated by taking $\textbf{z} - \textbf{h}$~(with $\textbf{h} =  \textbf{e}-\textbf{g}$) as the input to the HDD. Let $\textbf{f}\in \mathcal{F}(\textbf{e})$ be a flipping pattern such that $\lambda(\textbf{f}) = \min_{\textbf{h}\in \mathcal{F}(\textbf{e})} \lambda(\textbf{h})$. It suffices to prove that, for $W_H(\textbf{e}) > t_{\min}$ and $\textbf{e} = \textbf{f} + \textbf{g}$, we have $\textbf{g} \in \mathcal{G}(\textbf{f})$, i.~e., $|\mathcal{S}(\textbf{g})| = t_{\min}$, $\mathcal{S}(\textbf{f}) \bigcap \mathcal{S}(\textbf{g}) = \emptyset$ and $R_u(\textbf{f}) < R_{\ell}(\textbf{g}$). This can be proved by contradiction.

Suppose that $|\mathcal{S}(\textbf{g})| < t_{\min}$. Let $(i_1, \gamma_1)$ be a nonzero component of $\textbf{f}$. Then $\textbf{e}$ can also be generated by $\textbf{f} - (i_1, \gamma_1)$, whose soft weight is less than that of $\textbf{f}$, a contradiction to the definition of $\textbf{f}$.

Suppose that $\mathcal{S}(\textbf{f}) \bigcap \mathcal{S}(\textbf{g}) \neq \emptyset$. Let $i_1 \in \mathcal{S}(\textbf{f}) \bigcap \mathcal{S}(\textbf{g})$ and $(i_1, \gamma_1)$ be the nonzero component of $\textbf{f}$. Then $\textbf{e}$ can also be generated by $\textbf{f} - (i_1, \gamma_1)$, whose soft weight is less than that of $\textbf{f}$, a contradiction to the definition of $\textbf{f}$.

Obviously, $R_u(\textbf{f}) \neq R_{\ell}(\textbf{g})$ since their support sets have no common coordinates. Suppose that $R_u(\textbf{f}) > R_{\ell}(\textbf{g})$.  Let $(i_1, \gamma_1)$ be the atom with rank $R_u(\textbf{f})$ and $(i_2, \gamma_2)$ be the atom with rank $R_{\ell}(\textbf{g})$.  Then $\textbf{e}$ can also be generated by $\textbf{f} - (i_1, \gamma_1) + (i_2, \gamma_2)$, whose soft weight is less than that of $\textbf{f}$, a contradiction to the definition of $\textbf{f}$.
\end{IEEEproof}



%

%
%

\ifCLASSOPTIONcaptionsoff

\newpage
\fi



%

\bibliographystyle{IEEEtran}
\bibliography{ChaseGS}




%

%
%
%




\end{document}